\documentclass{Interspeech2024}

\interspeechcameraready

\usepackage{graphicx}
\usepackage{xcolor}
\usepackage[caption=false,font=footnotesize]{subfig}
\usepackage{url}

\usepackage{array, amsmath, amssymb, amsfonts, bm}
\usepackage{mathrsfs}
\usepackage{multirow, booktabs}
\usepackage{algorithm, algpseudocode}
\usepackage{xargs}

\usepackage{caption}
\usepackage{comment}
\usepackage{kantlipsum}
\usepackage{todonotes}
\usepackage{microtype}

\usepackage[
    backend=biber,
    style=ieee,
    citestyle=numeric-comp,
    minbibnames=1,
    maxbibnames=1,
    mincitenames=1,
    maxcitenames=1,
    doi=false,isbn=false,url=false,eprint=false]{biblatex}
\DeclareSourcemap{
\maps[datatype=bibtex, overwrite=true]{
\map{
    \step[fieldsource=booktitle,
    match=\regexp{.*International .*Workshop .*Machine.*Learning.*Signal.*Processing.*},
    replace={Proc. MLSP}]
    \step[fieldsource=booktitle,
        match=\regexp{.*Association.*for.*Computational.*Linguistics.*},
        replace={Proc. ACL}]
    \step[fieldsource=booktitle,
        match=\regexp{.*Interspeech.*},
        replace={Proc. Interspeech}]
    \step[fieldsource=booktitle,
        match=\regexp{.*INTERSPEECH.*},
        replace={Proc. Interspeech}]
    \step[fieldsource=journal,
        match=\regexp{.*INTERSPEECH.*},
        replace={Proc. Interspeech}]
    \step[fieldsource=booktitle,
        match=\regexp{.*ICASSP.*},
        replace={Proc. ICASSP}]
    \step[fieldsource=booktitle,
        match=\regexp{.*icassp_inpress.*},
        replace={Proc. ICASSP (in press)}]
    \step[fieldsource=booktitle,
        match=\regexp{.*Acoustics,.*Speech.*and.*Signal.*Processing.*},
        replace={Proc. ICASSP}]
    \step[fieldsource=booktitle,
        match=\regexp{.*European.*Signal.*Processing.*Conference.*},
        replace={Proc. EUSIPCO}]
    \step[fieldsource=booktitle,
        match=\regexp{.*International.*Conference.*on.*Learning.*Representations.*},
        replace={Proc. ICLR}]
    \step[fieldsource=booktitle,
        match=\regexp{.*International.*Conference.*on.*Computational.*Linguistics.*},
        replace={Proc. COLING}]
    \step[fieldsource=booktitle,
        match=\regexp{.*SIGdial.*Meeting.*on.*Discourse.*and.*Dialogue.*},
        replace={Proc. SIGDIAL}]
    \step[fieldsource=booktitle,
        match=\regexp{.*International.*Conference.*on.*Machine.*Learning.*},
        replace={Proc. ICML}]
    \step[fieldsource=booktitle,
        match=\regexp{.*North.*American.*Chapter.*of.*the.*Association.*for.*Computational.*Linguistics:.*Human.*Language.*Technologies.*},
        replace={Proc. NAACL}]
    \step[fieldsource=booktitle,
        match=\regexp{.*Empirical.*Methods.*in.*Natural.*Language.*Processing.*},
        replace={Proc. EMNLP}]
    \step[fieldsource=booktitle,
        match=\regexp{.*Association.*for.*Computational.*Linguistics.*},
        replace={Proc. ACL}]
    \step[fieldsource=booktitle,
        match=\regexp{.*Automatic.*Speech.*Recognition.*and.*Understanding.*},
        replace={Proc. ASRU}]
    \step[fieldsource=booktitle,
        match=\regexp{.*Spoken.*Language.*Technology.*},
        replace={Proc. SLT}]
    \step[fieldsource=booktitle,
        match=\regexp{.*Speech.*Synthesis.*Workshop.*},
        replace={Proc. SSW}]
    \step[fieldsource=booktitle,
        match=\regexp{.*workshop.*on.*speech.*synthesis.*},
        replace={Proc. SSW}]
    \step[fieldsource=booktitle,
        match=\regexp{.*Advances.*in.*neural.*information.*processing.*},
        replace={Proc. NeurIPS}]
    \step[fieldsource=booktitle,
        match=\regexp{.*Advances.*in.*Neural.*Information.*Processing.*},
        replace={Proc. NeurIPS}]
    \step[fieldsource=booktitle,
        match=\regexp{.*Workshop.*on.* Applications.* of.* Signal.*Processing.*to.*Audio.*and.*Acoustics.*},
        replace={Proc. WASPAA}]
    \step[fieldsource=publisher,
        match=\regexp{.+},
        replace={{}}]
    \step[fieldsource=month,
        match=\regexp{.+},
        replace={{}}]
    \step[fieldsource=location,
        match=\regexp{.+},
        replace={{}}]
    \step[fieldsource=address,
        match=\regexp{.+},
        replace={{}}]
    \step[fieldsource=organization,
        match=\regexp{.+},
        replace={{}}]
}}}

\defbibheading{bibliography}[\refname]{\section{#1}}
\title{Neural Blind Source Separation and Diarization \\ for Distant Speech Recognition}

\name[affiliation={1}]{Yoshiaki}{Bando}
\name[affiliation={1}]{Tomohiko}{Nakamura}
\name[affiliation={2}]{Shinji}{Watanabe}

\address{
  $^1$National Institute of Advanced Science and Technology (AIST), Japan\\
  $^2$Carnegie Mellon University, USA
}
\email{\{y.bando, tomohiko-nakamura\}@aist.go.jp, swatanab@andrew.cmu.edu}


\newcommand{\diag}{\mathrm{diag}}

\newcommand{\setR}{\mathbb{R}}
\newcommand{\setRp}{\mathbb{R}_+}
\newcommand{\setC}{\mathbb{C}}
\newcommand{\setSp}{\mathbb{S}_+}

\newcommand{\eye}{{\bf I}}

\newcommand{\adj}{\mathsf{H}}

\newcommand{\distbernoulli}[1]{\mathrm{Bernoulli}\left({#1}\right)}

\newcommand{\distnormal}[2]{\mathcal{N}\left({#1}, {#2}\right)}
\newcommand{\distcmpnormal}[2]{\mathcal{N}_{\mathbb{C}}\left({#1}, {#2}\right)}

\newcommand{\E}{\mathbb{E}}
\newcommand{\KL}{\mathcal{D}_\mathrm{KL}}

\NewDocumentCommand\newletter{m m o m m}{%
\NewDocumentCommand#1{s t@ o}{%
\IfBooleanTF{##1}{\mathbf{\MakeUppercase{#2}}\IfValueT{#3}{^{#3}}}{%
\IfBooleanTF{##2}{\mathbf{#2}\IfValueT{#3}{^{#3}}_{\IfValueTF{##3}{##3}{#5}}}{%
{#2}\IfValueT{#3}{^{#3}}_{\IfValueTF{##3}{##3}{#4}}%
}}}}

\NewDocumentCommand\newletterbm{m m o m m}{%
\NewDocumentCommand#1{s t@ o}{%
\IfBooleanTF{##1}{\bm{\MakeUppercase{#2}}\IfValueT{#3}{^{#3}}}{%
\IfBooleanTF{##2}{\bm{#2}\IfValueT{#3}{^{#3}}_{\IfValueTF{##3}{##3}{#5}}}{%
{#2}\IfValueT{#3}{^{#3}}_{\IfValueTF{##3}{##3}{#4}}%
}}}}

\newletter{\x}{x}{ftm}{ft}

\newletterbm{\psd}{\lambda}{nft}{ft}

\newcommand{\scm}[1][nf]{\mathbf{H}_{#1}}

\newcommand{\Q}[1][f]{\mathbf{Q}_{#1}}
\newletter{\q}{q}{fmm}{fm}
\newletter{\xt}{\tilde{x}}{ftm}{ft}
\newletter{\yt}{\tilde{y}}{nftm}{nft}

\newletter{\z}{z}{ntd}{nt}
\newletter{\zs}{z}[*]{ntd}{nt}
\newletter{\msk}{u}{nt}{t}

\newletter{\g}{w}{nfm}{nf}

\newcommand{\sv}[1][nf]{\mathbf{a}_{#1}}

\newcommand{\dec}[1][\theta,f]{g_{#1}}

\newletter{\src}{s}{nft}{nf}

\newcommand{\elbo}{\mathcal{L}}

\begin{document}
\setlength\abovedisplayskip{1.7mm}
\setlength\belowdisplayskip{1.7mm}

\maketitle

\begin{abstract}
This paper presents a neural method for distant speech recognition (DSR) that jointly separates and diarizes speech mixtures without supervision by isolated signals. A standard separation method for multi-talker DSR is a statistical multichannel method called guided source separation (GSS). While GSS does not require signal-level supervision, it relies on speaker diarization results to handle unknown numbers of active speakers. To overcome this limitation, we introduce and train a neural inference model in a weakly-supervised manner, employing the objective function of a statistical separation method. This training requires only multichannel mixtures and their temporal annotations of speaker activities. In contrast to GSS, the trained model can jointly separate and diarize speech mixtures without any auxiliary information. The experiments with the AMI corpus show that our method outperforms GSS with oracle diarization results regarding word error rates. The code is available online.
\end{abstract}

\section{Introduction}  
Speech separation and enhancement are essential functions for multi-talker distant speech recognition (DSR) from noisy mixture recordings~\cite{kanda2023vararray,tripathi2020end,von2020multi,cornell23_chime,watanabe2020chime,wang2023ustc,medennikov2020stc}.
As represented by teleconference systems and conversational robots, speech signals are often recorded by microphones located at distance from the speakers. 
Such recordings are thus often contaminated by other speakers' utterances and environmental noise, which significantly degrade the recognition performance~\cite{kanda2023vararray,tripathi2020end,von2020multi}.
This calls for speech separation methods that can handle unknown and dynamically changing numbers of active speakers in diverse noisy environments.

Blind source separation (BSS) has widely been utilized in DSR because sufficient and matched-domain training data of isolated signals are often unavailable for conversational recordings~\cite{boeddecker2018frontenda,raj2022gpu,shimada2019unsupervised,drude2019unsupervised2,wang2024unssor}.
The guided source separation (GSS)~\cite{boeddecker2018frontenda,raj2022gpu}, for example, estimates time-frequency (TF) masks for active speakers based on a complex angular central Gaussian mixture model (cACGMM)~\cite{ito2016complex}.
One drawback of the cACGMM is its performance limitation due to the sparse assumption that each TF bin contains only one source.
To overcome this limitation, full-rank spatial covariance analysis (FCA)~\cite{duong2010underdetermined,sawada2020experimental} and its extensions~\cite{shimada2019unsupervised,sekiguchi2020fast} have been investigated by assuming each TF bin as the sum of all the sources.
FCA has further been extended for unsupervised training of a neural separation model by maximizing its log-marginal likelihood~\cite{bando2022weakly,bando2021neural}.
This method, called neural FCA, was reported to outperform GSS and existing BSS methods.

Most of the BSS methods, including GSS and neural FCA, assume that the number of sound sources is known in advance.
Performing BSS with an incorrect number of sources can cause the under- or over-separation problem.
GSS~\cite{boeddecker2018frontenda} solves this problem by masking source activities with speaker diarization results provided in advance.
This approach, however, may often fall into sub-optimal solutions because the speaker diarization and separation are in a chicken-and-egg relationship~\cite{medennikov2020stc,wang2023ustc}.
A supervised neural model, called end-to-end neural diarization and source separation (EEND-SS)~\cite{maiti2023eend}, thus has been proposed to jointly separate and diarize speech mixtures in a unified network architecture.
This method, however, requires oracle isolated signals to train source separation, which constrains its applicability to multi-talker DSR systems.
For example, in the CHiME-7 DSR challenge~\cite{cornell23_chime}, no participant was able to make supervised neural separation to work because of the domain mismatch.  %

\begin{figure}[t]
  \setlength\abovecaptionskip{1.5mm}
  \setlength\belowcaptionskip{0mm}
  \centering
  \includegraphics[width=\hsize]{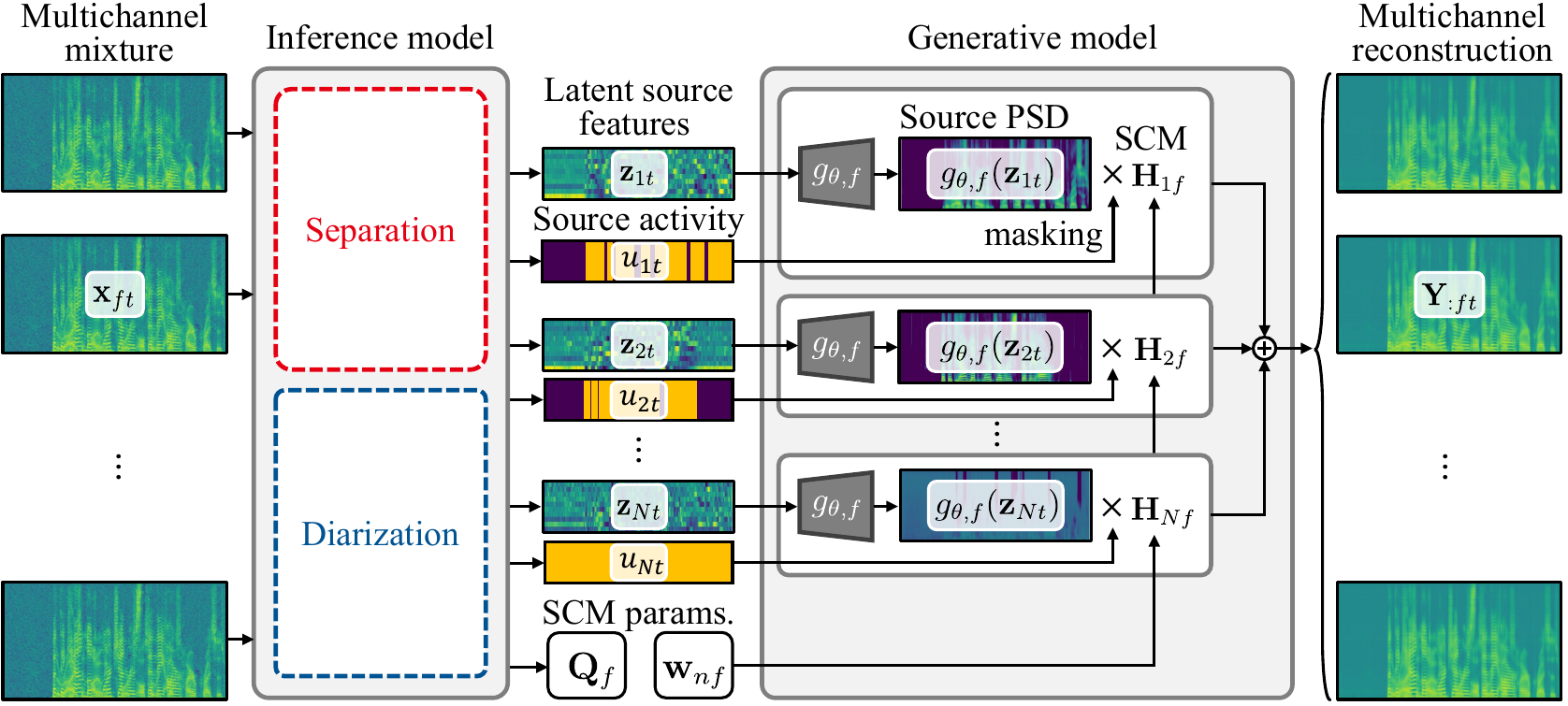}
  \caption{The overview of our joint separation and diarization.}
  \label{fig:overview}
\end{figure}

In this paper, we propose a weakly-supervised method to perform joint speech separation and diarization by a multitask learning of unsupervised separation and supervised diarization (Fig.~\ref{fig:overview}).
We take advantage of the BSS techniques that separate speech signals utilizing the spatial information of multichannel mixtures.
Specifically, the unsupervised separation is trained based on the objective function of the neural FCA.
The supervised diarization is, on the other hand, trained to minimize the binary cross entropy as in the EEND-SS.
Since there is a permutation ambiguity between the estimated sources and oracle activations, we solve this problem by utilizing permutation invariant training (PIT)~\cite{maiti2023eend}.
Once the network is trained, it can perform its inference only with multichannel mixtures unlike the conventional BSS methods.

The main contribution of this study is to solve speech separation and diarization by taking full advantage of the statistical and neural frameworks.
While the speech separation has been actively solved by using the unsupervised BSS techniques, the diarization has been solved by the supervised neural training.
We combine these statistical and neural frameworks into a unified inference model with neural BSS training.
We demonstrate that such a compound architecture can be successfully trained from real audio mixtures of the AMI corpus.
The experimental results show that our method outperforms GSS with the oracle speaker activities regarding word error rates (WERs).
The diarization error rate (DER), in addition, is significantly improved from that for signals separated by the original neural FCA.

\section{Background}

This section describes the formulation of BSS and introduces the neural FCA, which is extended to the proposed joint source separation and diarization.

\subsection{Blind source separation}
The typical BSS method assumes that an $M$-channel mixture signal $\x@ \in \setC^M$ is a sum of $N$ source signals $\src \in \setC$ in the short-time Fourier transform (STFT) domain:
\begin{align}
  \x@ = \sum_{n=1}^N \sv \src, \label{eq:problem}
\end{align}  %
where $\sv \in \setC^M$ is the steering vector for source $n$, and $t=1,\ldots, T$ and $f=1,\ldots, F$ represent the time and frequency indices, respectively.
Each source signal $\src$ is assumed to follow a complex zero-mean Gaussian distribution with a power spectrum density (PSD) $\psd \in \setRp$ as follows:
\begin{align}
  \src \sim \distcmpnormal{0}{\psd}. \label{eq:src}
\end{align}
By marginalizing $\src$ from Eqs.~\eqref{eq:problem} and \eqref{eq:src}, the following multivariate Gaussian likelihood is obtained:
\begin{align}
  \x@ \sim \distcmpnormal{\bm{0}}{\sum_{n=1}^N \psd \scm}, \label{eq:lgm}
\end{align}
where $\scm = \sv \sv^\adj \in \setSp^{M\times M}$ is the spatial covariance matrix (SCM) of source $n$ at frequency $f$.
This model is known to be robust against diffuse noise and small source movements by allowing the full-rankness of the SCMs~\cite{duong2010underdetermined,sawada2013multichannel}.

Since the inference of the full-rank SCMs is computationally demanding, its reduction has been investigated~\cite{duong2010underdetermined,sawada2020experimental,ito2016complex,sekiguchi2020fast,ito2019fastmnmf}.
The cACGMM and GSS, for example, reduce the computational cost by assuming that each TF bin of Eq.~\eqref{eq:lgm} has only one of all the sources\footnote{Eq.~\eqref{eq:lgm} with this assumption and the cACGMM are identical in their maximum likelihood estimation~\cite{ito2016complex}.}.
In contrast, joint diagonalization (JD)~\cite{sekiguchi2020fast,ito2019fastmnmf} was proposed to reduce the cost while maintaining that each TF bin is the sum of all the sources.
The JD assumes that the SCM $\scm$ is diagonalized by $\Q \in \setC^{M\times M}$ common for all the sources:
\begin{align}
  \scm = \Q^{-1} \diag(\g@) \Q^{-\adj}, \label{eq:jd}
\end{align}
where $\g@ \in \setRp^M$ is a diagonal coefficient for source $n$.
The parameters of $\Q[] \triangleq \{ \Q \}_{f=1}^F$ and $\g* \triangleq \{\g@\}_{n,f=1}^{N,F}$ are efficiently estimated by the iterative source steering (ISS) algorithm~\cite{scheibler2020fast,sekiguchi2020fast}.
The separation performance with the JD SCMs was reported to be comparable to that of the full-rank SCMs~\cite{sekiguchi2020fast}.

Another important factor for BSS is how to represent the source PSDs $\psd$ precisely.
A promising approach is deep spectral modeling based on variational autoencoders (VAEs)~\cite{leglaive2020recurrent,bando2021neural,Li2023,kingma2013auto}.
This model typically introduces $D$-dimensional latent source features $\z@ \in \setR^D$ to generate the PSD $\psd$ with a deep neural network (DNN) $\dec: \setR^D \rightarrow \setRp$ as follows:
\begin{align}
  \psd = \dec(\z@), \label{eq:dsp}
\end{align}
where $\theta$ represents the model parameters of $\dec$.
The latent features $\z@$ are supposed to represent spectral characteristics (e.g., pitches and envelopes).
This model is trained as a decoder of a VAE for isolated signals by assuming that $\z@$ follows the standard Gaussian distribution:
\begin{align}
  \z@ \sim \distnormal{\bm{0}}{\eye}. \label{eq:pz}
\end{align}
After training, the sources are separated from a mixture by estimating $\z@$ to maximize the observation likelihood of Eq.~\eqref{eq:lgm}.

\subsection{Neural full-rank spatial covariance analysis}
Neural FCA has been proposed to train the deep spectral model and its inference (separation) model only from multichannel mixtures~\cite{bando2023neural}.
The inference model $h_\phi$ is introduced to estimate latent source features $\z*\triangleq \{\z@\}_{n,t=1}^{N,T}$ in Eq.~\eqref{eq:dsp} from a multichannel mixture $\x* \triangleq \{\x@\}_{f,t=1}^{F,T}$ in Eq.~\eqref{eq:problem} as a posterior distribution $q_\phi(\z* \mid \x*)$:
\begin{align}
  q_\phi(\z* \mid \x*) \leftarrow h_\phi(\x*),
\end{align}
where $\phi$ represents the network parameters.
The generative and inference models are jointly trained to maximize an evidence lower bound (ELBO) derived from Eqs.~\eqref{eq:lgm}, \eqref{eq:dsp}, and \eqref{eq:pz}:
\begin{align}
  \elbo^\mathrm{(sep)} &= \E_{q_\phi}[\log p_\theta(\x* \mid \z*, \scm[])] - \nonumber \\[-1mm]
  &\hspace{25mm}\KL[q_\phi(\z*\mid \x*) \mid p(\z*)], \label{eq:elbo}
\end{align}  %
where $\E_{q_\phi}[\cdot]$ is the expectation by $q_\phi$, and $\KL[p\mid q]$ represents the Kullback-Leibler (KL) divergence between $p$ and $q$.
The SCM $\scm$ is obtained by an expectation-maximization (EM) algorithm~\cite{duong2010underdetermined,bando2021neural}.
This method can be considered as a large VAE for multichannel mixture signals, in which the decoder is defined by the multichannel generative model of Eq.~\eqref{eq:lgm}.

One relevant work to our study is weakly-supervised (WS) neural FCA for a front-end system of DSR~\cite{bando2022weakly}.
To handle the dynamically changing number of active speakers, this method introduces the source activity mask $\msk \in \{0, 1\}$ in a similar way to GSS.
The inference model $h_{\phi}$ is extended from Eq.~(7) to utilize the mask $\msk$ as a condition and estimate the correct number of latent source features:  %
\begin{align}
  q_\phi(\z* \mid \x*, \msk*) \leftarrow h_\phi(\x*, \msk*).
\end{align}
While this method was reported to outperform the GSS in the CHiME-6 corpus, it assumes the mask $\msk$ to be known in advance.
Our study aims to remove this limitation to perform the inference only with multichannel mixture signals.

\begin{figure}[t]
  \setlength\abovecaptionskip{0.0mm}
  \setlength\belowcaptionskip{-3mm}
  \centering
  \includegraphics[width=0.8\hsize]{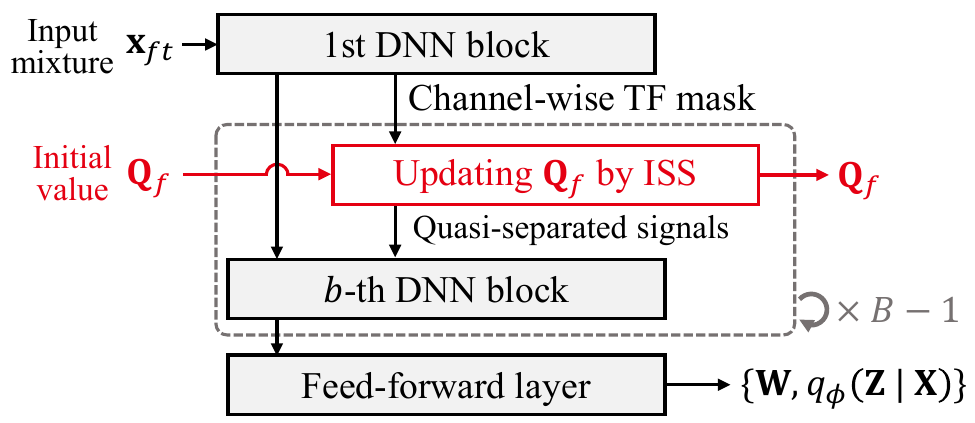}
  \caption{The inference model of neural FastFCA, which employs the hybrid architecture proposed in \cite{Scheibler2021}.}
  \label{fig:dnn-iva}
\end{figure}

Another relevant work called neural FastFCA~\cite{bando2023neural} introduces the JD (Eq.~\eqref{eq:jd}) to reduce the computational cost.
The inference model $h_\phi$ is extended from Eq.~(7) to estimate the diagonalizer $\Q[]$ and diagonal elements $\g*$ in addition to $q_\phi(\z* \mid \x*)$:
\begin{align}
  \left\{ \Q[], \g*, q_\phi(\z* \mid \x*)  \right\} \leftarrow h_\phi(\x*). \label{eq:inf_fastfca}
\end{align}
As illustrated in Fig.~\ref{fig:dnn-iva}, this inference model is designed to efficiently estimate these parameters by alternately performing the ISS algorithm and DNN inference~\cite{Scheibler2021}.
The neural FastFCA was reported to significantly reduce the inference time from that of the neural FCA without performance degradation~\cite{bando2023neural}.

\section{Joint Speech Separation and Diarization Based on Neural FastFCA}
\setlength\abovedisplayskip{1.1mm}
\setlength\belowdisplayskip{1.1mm}
The proposed method performs joint separation and diarization by taking a full advantage of neural FastFCA.  %
Our method, called neural FCA with speaker activity (FCASA), is an extension of the neural FastFCA in Eq.~\eqref{eq:inf_fastfca} to estimate the time-varying number of active speakers in addition to the separation parameters.

\subsection{Generative model of multichannel mixture signals}
To handle the unknown and time-varying number of active speakers, we assume $N$ in Eq.~\eqref{eq:problem} as a possible maximum number of sources and introduce the speaker activity $\msk \in \{0, 1\}$ as:
\begin{align}
  \x@ = \sum_{n=1}^N \msk \sv \src.
\end{align}
The mask is estimated by an inference model (detailed in the next section) unlike the existing GSS and WS neural FCA.
As in the neural FastFCA, we also introduce the JD SCMs (Eq.~\eqref{eq:jd}). The resulting likelihood function is derived as follows:
\begin{align}
  \x@ \sim \distcmpnormal{\bm{0}}{\Q^{-1}\left\{ \sum_{n=1}^N \diag\left( \yt@ \right) \right\}\Q^{-\adj}},
\end{align}
where $\yt@ \triangleq \msk \dec(\z@) \g@ \in \setRp^M$ is the source PSD in the diagonalized space $\Q\x@$.

\subsection{Inference model}

We design an inference model $h_\phi$ to jointly separate and diarize an input multichannel mixture.
Specifically, this model $h_\phi$ outputs the posterior distributions of the source features $\z*$ and the speaker activity $\msk$ as well as the JD parameters of $\Q[]$ and $\g*$:
\begin{align}
  \left\{ \Q[], \g*, q_\phi(\z* \mid \x*), q_\phi(\msk* \mid \x*) \right\} \leftarrow h_\phi(\x*). \label{eq:inference}
\end{align}
The posterior distributions $q_\phi$ are defined by network outputs $\mu_{\phi,ntd} \in \setR$, $\sigma^2_{\phi,ntd} \in \setRp$, and $\eta_{\phi,nt} \in [0, 1]$ as follows:
\begin{align}
  q_\phi(\z* \mid \x*) &\triangleq \prod_{n,t,d=1}^{N,T,D} \distnormal{\z \bigm| \mu_{\phi,ntd}}{\sigma^2_{\phi,ntd}}, \\[-1mm]
  q_\phi(\msk* \mid \x*) &\triangleq \prod_{n,t=1}^{N,T} \distbernoulli{\msk \bigm| \eta_{\phi,nt}},
\end{align}
At the inference phase, the source signals are separated by a multichannel Wiener filter~\cite{sekiguchi2020fast,bando2023neural} using  $\Q$, $\g@$, and $\mu_{\phi,ntd}$.
The diarization results are obtained as temporal speech activities by thresholding $\eta_{\phi,nt}$.
Note that we have not explicitly defined the prior distribution for $\msk$ because its posterior is trained as an empirical distribution in a supervised manner.

While the inference model (Fig.~\ref{fig:dnn-iva}) of the original neural FastFCA utilizes a UNet-like architecture~\cite{tzinis2020sudo}, we utilize the resource-efficient (RE)-SepFormer~\cite{subakan2022resource} to handle the long-term dependencies of speech activities.
Since the original RE-SepFormer is designed for monaural separation, we introduce the transform-average-concatenate (TAC)~\cite{luo2020end} models for the inter-channel communication of the channel-wise inference.
The TAC modules are inserted to the middles of RE-SepFormer blocks.  %

\subsection{Training without isolated source signals}
\setlength\abovedisplayskip{1.3mm}
\setlength\belowdisplayskip{1.3mm}
We train the inference model $h_\phi$ (Eq.~\eqref{eq:inference}) and generative model $g_\theta$ (Eq.~\eqref{eq:dsp}) in a multi-task learning of unsupervised separation and supervised diarization.
We use multichannel mixture signals $\x*$ and their temporal annotations of speaker activities $\msk*$ for the training data.
The objective function to be maximized is the weighted sum of the functions for unsupervised separation $\elbo^\mathrm{(sep)}$ and supervised diarization $\elbo^\mathrm{(diar)}$ as follows:
\begin{align}
  \elbo &= \frac{1}{TF} \elbo^\mathrm{(sep)} + \gamma \frac{1}{TN}\elbo^\mathrm{(diar)}, \label{eq:objective}
\end{align}
where $\gamma \in \setRp$ is a scaling hyperparameter.

The separation term $\elbo^\mathrm{(sep)}$ trains the estimation of $\Q[]$, $\g*$, and $q_\phi(\z* \mid \x*)$ by using the ELBO of the neural FastFCA:
\begin{align}
  \elbo^\mathrm{(sep)} &= \E_{q_\theta}[\log p_\theta(\x* \mid \msk*, \z*, \g*, \Q[])] \nonumber \\
  &\hspace{25mm}- \KL[q_\phi(\z* \mid \x*) \mid p(\z*)].
\end{align}
The first term of this ELBO is calculated as follows:
{
\setlength\abovedisplayskip{0.5mm}
\setlength\belowdisplayskip{0.5mm}
\begin{align}
  \E_{p_\theta}[\log p_\theta\left( \x* \mid \msk*, \z*, \g*, \Q[] \right)] &\approx T \sum_{f=1}^F \log \left|\Q\Q^\adj\right| \nonumber \\[-2.5mm]
  & \hspace{-20mm}- \sum_{f,t,m=1}^{F, T, M} \left\{ \log \yt[:ftm] + \frac{|\xt|^2}{\yt[:ftm]} \right\},
\end{align}}%
where $\xt@ \triangleq [\xt[ft1],\ldots, \xt[ftM]]^T = \Q \x@ \in \setC^M$ represents the diagonalized (quasi-separated) observation, and $\yt[:ftm] \triangleq\sum_n \yt$ is calculated from the sample of $q_\phi(\z* \mid \x*)$.
We use the oracle speaker activities for the mask $\msk$ as teacher forcing.
The maximization of this objective function is equivalent to the maximization of the log-marginal likelihood $\log p_\theta(\x* \mid \msk*, \g*, \Q[])$.
For $q_\phi(\z* \mid \x*)$, it also corresponds to the minimization of the KL divergence between the network estimate and the oracle posterior $\KL[q_\phi(\z* \mid \x*)\mid p_\theta(\z*\mid\x*,\Q[], \g*, \msk*)]$.

The diarization term $\elbo^\mathrm{(diar)}$, on the other hand, directly maximizes the log-posterior as supervised training:
\begin{align}
  \elbo^\mathrm{(diar)} &\triangleq \log q(\msk* \mid \x*) = -\mathrm{BCE}\left[\msk \mid \eta_{\phi,nt}\right], 
\end{align}
where $\mathrm{BCE}\left[\cdot \mid \cdot\right]$ represents the binary cross entropy.
The maximization of this objective corresponds to the minimization of the KL divergence between the empirical posterior distribution and the network estimate $\KL[p_\mathrm{data}(\msk*\mid\x*) \mid q_\phi(\msk* \mid \x*)]$.
Source indices $\{1,\ldots, N\}$ are permuted to minimize $\elbo$ as PIT.

\section{Experimental Evaluation}
We evaluated our neural FCASA by using a real meeting dataset called the AMI corpus~\cite{kraaij2005ami}.
The training and inference scripts with a pre-trained model are available in \url{https://ybando.jp/projects/neural-fcasa/}.

\subsection{Dataset}
The AMI corpus contains approximately 100 hours of English meeting recordings.
The recording was performed in three meeting rooms at different institutes in European countries, and there were three to five participants in each meeting.
We utilized the audio signals recorded by a microphone array placed on the meeting table in each room.
The array has a circular shape with eight microphones ($M=8$) and a radium of 10\,cm.
We used the official split of training, development, and evaluation subsets having 80.7 hours, 9.7 hours, and 9.1 hours, respectively.
They were recorded in 48\,kHz and resampled to 16\,kHz~\cite{kraaij2005ami}.
All the recordings were dereverberated in advance by using the weighted prediction error (WPE) method~\cite{yoshioka2012generalization}.

\subsection{Experimental configurations}
The network architectures of the proposed neural FCASA were determined experimentally as follows.
The encoder consisted of eight RE-SepFormer blocks~\cite{subakan2022resource}, with ISS blocks~\cite{Scheibler2021} inserted twice between them.
The RE-SepFormer blocks consisted of the Transformer encoder layers having 256 latent units with a feed-forward dimension of 1024 and eight multi-head attentions.
The decoder consisted of six linear layers, each having 256 channels with residual connections and parametric rectified linear units (PReLUs).
The nonnegativity of the decoder outputs was obtained by the softplus activation.

The networks were trained for 200 epochs by an AdamW optimizer with the learning rate of $1.0\times 10^{-4}$ and the weight decay of $1.0\times 10^{-5}$.
The spectrograms were obtained by the STFT with the window size of $512$ samples and the hop length of $160$ samples.
The maximum number of sources $N$ was set to $6$ by assuming five speakers ($n=1,\ldots, N-1$) at maximum and one noise source ($n=N$).
The dimension for the latent source features $D$ was set to 64.
Following \cite{bando2022weakly}, to prevent the noise source ($n=N$) from representing speaker utterances, we set the dimension $D$ for noise to 10.
The speaker activations $\msk$ were obtained by using the oracle diarization results, while that for noise was set to always active.
$\gamma$ in Eq.~\eqref{eq:objective} was set to $1.0$.
The training data was split into 20-second clips and fed to the training pipeline with their random crops of 10 seconds.
The batch size was set to 128.
These hyperparameters and architectures were determined experimentally by using the validation set.

\begin{table}[t]
  \centering
  \setlength\abovecaptionskip{1mm}
  \setlength\tabcolsep{0.52mm}
  \newcommand{\mr}[1]{\multirow{2}{*}{#1}}
  \newcommand{\mc}[2]{\multicolumn{4}{#1}{#2}}
  
  \fontdimen16\textfont2=0pt
  \fontdimen17\textfont2=0pt
  \newcommand{\ci}[3]{$#1$\hspace{0.5mm}\textcolor{gray}{$^{\scriptscriptstyle #3}_{\hspace{-0.1mm}\smash{\raisebox{-0.1mm}{$\scriptscriptstyle#2$}}}$}}
  \caption{SCAs, DERs, and WERs with their 95\% confidence intervals.
  ``Diar. free" means that the separation method is free from the diarization results.
  The non-free methods (i.e., GSS and WS Neural FCA) used the oracle diarization results.}
  \label{tab:scores}
  \small
  \begin{tabular}{l|c|cc|cc}
    \toprule
    \mr{Method} & Diar. & \mr{SCA$^\uparrow$} & \mr{DER$^\downarrow$} & WER$^\downarrow$ & WER$^\downarrow$ \\
                & free  &    &  & (AMI)   & (OWSM) \\
    \midrule
    Headset mic.  &           -- &   -- &   -- & \ci{18.3}{-0.4}{+0.4} & \ci{19.2}{-1.6}{+1.8} \\
    Array mic.    &           -- &   -- &   -- & \ci{59.7}{-0.7}{+0.7} & \ci{52.0}{-3.1}{+3.4} \\
    \midrule
    GSS           &              &   -- &   -- & \ci{36.3}{-0.6}{+0.6} & \ci{28.7}{-1.5}{+1.7} \\
    cACGMM        & $\checkmark$ &   -- &   -- & \ci{44.9}{-0.6}{+0.6} & \ci{34.9}{-1.9}{+2.2} \\
    FastMNMF2     & $\checkmark$ &   -- &   -- & \ci{42.6}{-0.7}{+0.7} & \ci{33.7}{-1.9}{+2.4} \\
    \midrule
    WS Neural FCA &              &   -- &   -- & \ci{32.8}{-0.6}{+0.6} & \ci{28.2}{-1.8}{+2.2} \\
    Neural FCA    & $\checkmark$ & \ci{14.8}{-1.1}{+1.2} & \ci{82.4}{-1.5}{+1.6} & \ci{33.3}{-0.6}{+0.6} & \ci{28.5}{-1.7}{+2.1} \\
    \midrule
    Neural FCASA  & $\checkmark$ & \ci{75.6}{-1.5}{+1.5} & \ci{14.1}{-0.5}{+0.5} & \ci{33.2}{-0.6}{+0.6} & \ci{27.0}{-1.6}{+1.9} \\
    \bottomrule 
  \end{tabular}
  \vspace{-3mm}
\end{table}

Our method was evaluated in the WERs, DERs, and source counting accuracies (SCAs).
We obtained WERs by utilizing two pre-trained ASR models publicly available for ESPnet~\cite{watanabe2018espnet}.
One is a standard Transformer-based model trained only on the headset recordings in the AMI corpus\footnote{\url{https://zenodo.org/records/4615756}}.
The other is a large-scale pre-trained model called the Open Whisper-style Speech Model (OWSM) v3.1 Medium~\cite{peng2024owsm}.
This model is based on the E-Branchformer~\cite{kim2023branchformer} and was trained on 180k hours of public speech data, including the AMI corpus.
The WER was calculated for crops of mixture signals, each having a minimum length of 10 seconds and a target utterance at its center.
We performed our method on them and extracted the target by aligning the estimated and oracle diarization results.
White noise was added to the estimates with signal-to-noise ratio of 40\,dB for alleviating their distortions.
Note that, while the ASR model in \cite{raj2022gpu} was trained by using separation results, we did not because we focus on the frontend performance.
We evaluated the DERs and SCAs for 10-second clips obtained by splitting the whole mixture recordings.
To stabilize the diarization results, the outputs $\eta_{\phi,nt}$ were smoothed by a median filter with a filer size of 11 frames.

We compared our method with the following existing BSS methods.
For statistical methods, we evaluated GSS~\cite{boeddecker2018frontenda}, cACGMM~\cite{ito2016complex}, and fast multichannel nonnegative matrix factorization 2 (FastMNMF2)~\cite{sekiguchi2020fast}.
FastMNMF2 is a state-of-the-art BSS method that utilizes JD SCMs.
We also evaluated the original neural FCA and WS neural FCA.
For a fair comparison, we implemented them with the JD SCMs and the same network architecture as the proposed method\footnote{They are FastFCA in precise, but we call them FCA for simplicity.}.
Since the cACGMM, FastMNMF2, and neural FCA are completely blind, we set the number of sources $N$ to 6 and aligned the source signals to the target utterances by using the oracle diarization results.
The SCA and DER were calculated for neural FCA by performing voice activity detection (VAD) to the separated signals.
We utilized a public VAD model pre-trained on the AMI corpus~\cite{bredin2023pyannote}.

\begin{figure}[t]
  \setlength\abovecaptionskip{1.0mm}
  \setlength\belowcaptionskip{-1mm}
  \captionsetup[subfloat]{captionskip=0mm}
  \vspace{-3mm}
  \subfloat[Neural FCA]{\includegraphics[width=0.49\hsize]{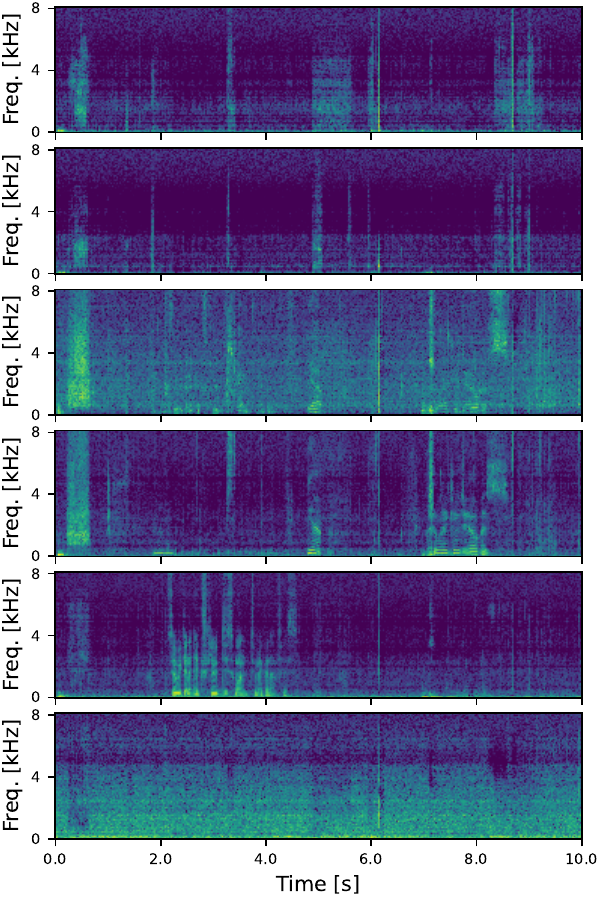}}
  \hfill
  \subfloat[Neural FCASA]{\includegraphics[width=0.49\hsize]{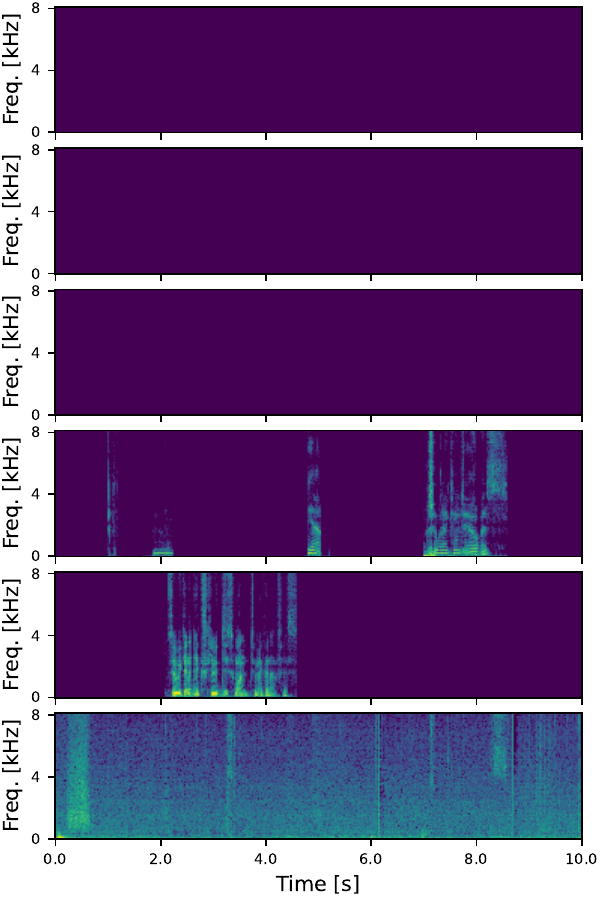}}
  \captionof{figure}{Examples of separation results by Neural FCA and Neural FCASA. Neural FCA over-separated one source into third and fourth results.}
  \label{fig:spec}
\end{figure}

\subsection{Experimental results}
The WERs, DERs, and SCAs are summarized in Table~\ref{tab:scores}.
We first see that the WERs of the statistical BSS methods (cACGMM and FastMNMF2) significantly deteriorated from that of GSS with the oracle speaker activities.
The inaccurate number of sources caused the degradation of separation performance in the classical methods.
The WS neural FCA improved WERs from GSS for the AMI-specific decoder, and the WS and original neural FCA performed comparably.
The DER and SCA of the neural FCA were, however, extremely poor.
As shown in Fig.~\ref{fig:spec}-(a), it tended to over-separate one utterance into multiple sources, which would degrade the DER and SCA.
In contrast, our neural FCASA maintained better performance in all the SCA, DER, and WERs.
Thanks to the supervised diarization training, the speech utterances were successfully aggregated as shown in Fig.~\ref{fig:spec}-(b).

\section{Conclusion}
We presented neural FCASA, which performs joint speech separation and diarization for DSR.
By combining the unsupervised BSS and supervised diarization techniques, our method trains an inference model without supervision by isolated signals.
Once trained, the model can be used to jointly separate and diarize speech mixtures without auxiliary information.
The experimental results with the AMI corpus show that our method outperformed GSS with oracle diarization results in WERs.
The future work includes extending our method to a continuous method as in \cite{kanda2023vararray} for sequentially separating and diarizing mixture signals.

\section{Acknowledgement}
This study was supported by the BRIDGE program of the Cabinet Office, Government of Japan.
We thank Dr. Samuele Cornell and Dr. Yoshiki Masuyama for their valuable discussion.

\printbibliography %

\end{document}